\documentstyle{plenum} 

\title{ULTRAHIGH-ENERGY NEUTRINO INTERACTIONS\\ 
AND NEUTRINO TELESCOPE EVENT RATES
\footnote{\rm\baselineskip=11pt
Talk presented by I. Sarcevic at {\it International Conference on 
Neutrino Mass, Dark Matter and Gravitational Waves}, 
Miami Beach, 
Florida, January 25 - 28, 1996.}}
\author
{Raj Gandhi,$^{1}$ Chris Quigg,$^{2}$ M. H. Reno,$^{3}$ 
and Ina 
Sarcevic$^{4}$\\ \\
${^{1}}$Mehta Research Institute, \\ 
10, Kasturba Gandhi Marg, Allahabad 211002, India \\
${^{2}}$Fermi National 
Accelerator Laboratory, \\
Batavia, IL 60510 USA \\
${^{3}}$Department of Physics and Astronomy, \\
University of Iowa,
Iowa City, IA 52242 USA \\
${^{4}}$Department of Physics, \\
University of Arizona,
Tucson, AZ 85721 USA}
\begin{document}
\maketitle                 
\vglue 0.8cm
\begin{center}
{\bf ABSTRACT}
\end{center}
\vglue 0.3cm
{\rightskip=3pc
 \leftskip=3pc
 \rm\baselineskip=12pt
 \noindent
We present results for 
neutrino-nucleon cross sections for energies up to 
$10^{21}$eV, of relevance to the detection of ultrahigh energy 
galactic and extragalactic neutrinos.  At the highest energies, 
our results are about 
$2.4$ times larger than previous estimates.  Using these new 
cross sections, we predict neutrino telescope event rates 
for the upward moving muons initiated by the neutrino interactions 
in the Earth and for the contained-vertex events in the PeV 
range 
due to neutrino-electron interactions.  We show that future 
neutrino detectors, such as AMANDA, BAIKAL, DUMAND and NESTOR 
have a very good chance of detecting neutrinos which originate in 
the Active Galactic Nuclei.  
}
\bigskip

\noindent{\bf{INTRODUCTION}}
\medskip

The Active Galactic Nuclei (AGN), with typical luminosities in the 
range $10^{42}$ to $10^{48}$ erg/s, are believed to be the most powerful 
individual sources of radiation in the Universe.  These extragalactic 
point sources 
are also considered as prodigious particle accelerators 
presumably powered by the gravitational energy of matter spiraling in 
to a supermassive black hole, though the mechanism responsible for the 
conversion of gravitational energy to luminous energy is not presently 
understood.  Recent detection of energetic photons 
($E_\gamma \sim 100$ MeV) 
from about 40 AGNs by the EGRET collaboration \cite{EGRET} and 
of TeV photons from 
Mkn 421, Mkn 501 \cite{AGN} and most recently from 1ES2344+514 by the 
Whipple 
collaboration \cite{Weekes} 
have created new excitement in the field of high-energy gamma-ray physics.  
If the observed photons are
decay products of $\pi^0$s produced in hadronic interactions in
the disk surrounding the AGN, then AGNs are also powerful sources of
ultrahigh-energy (UHE) neutrinos \cite{Ghs}.  
Unlike photons, which are absorbed by a few hundred gm/cm$^2$ of 
material, TeV neutrinos have interaction lengths
on the order of $250$ kt/cm$^2$ and 
thus can provide 
a direct window to the most 
energetic processes in the universe.

The advantage of the long interaction length translates to a challenge
in the detection of neutrinos. Interaction rates increase with energy,
but the fluxes of UHE neutrinos are steeply falling functions of
neutrino energy. Cerenkov detection of muons from
interactions of muon neutrinos in the
rock or ice surrounding the detector is feasible\cite{Detect}. More difficult
is the detection of charged-current interactions of electron neutrinos.
Large-area air shower arrays or large volume underground
detectors
may be adequate for the detection of 
electron neutrinos, especially near the $W$-boson resonance in
$\bar{\nu}_e e$ collisions.
Theoretical calculations of the neutrino-nucleon and neutrino-electron
cross sections are instrumental in evaluating event rates for neutrino
telescopes.

Here we present results
\cite{Gqrs}  
for charged current and neutral current 
cross sections for 
energies up to $10^{21}$ eV 
obtained 
using new parton distributions  
measured in 
$ep$
collisions at HERA \cite{He}.  
We also discuss how detection 
of UHE neutrinos depends on these cross sections and on the 
neutrino fluxes from UHE neutrino sources.  
Event rates for muon neutrino conversions to muons are compared with
earlier results based on older parton distribution functions \cite{Rq}.
We also present results for contained
events with higher threshold energies. 

\bigskip
\noindent{\bf SOURCES OF UHE NEUTRINOS}
\medskip

A variety of sources may contribute to the neutrino flux
at the surface of the Earth. Three types of sources are discussed
here: atmospheric neutrinos from cosmic-ray interactions in the
atmosphere, neutrinos from active galactic nuclei, 
and cosmic neutrinos from
extragalactic cosmic ray interactions with the microwave background
radiation. Model predictions for neutrino fluxes from these three types of
sources are shown in Figure 1.
Atmospheric neutrinos\cite{volkova} (ATM), while interesting in their own
right, mask extraterrestrial sources for $E_\nu<1$ TeV. Consequently,
we restrict our discussion to neutrino energies above 1 TeV.

\begin{figure}[htbp]
\vspace*{10.0 cm}
\caption{Muon neutrino plus antineutrino fluxes at the Earth's surface:
angle-averaged flux from atmospheric neutrinos (ATM),
diffuse flux from active galactic nuclei 
(AGN-NMB, AGN-SP and AGN-SS) and cosmic
neutrinos (CR-2 and CR-4). The Fr\'{e}jus upper limit$^{14}$
on a neutrino flux in 
excess of atmospheric neutrino flux is indicated at 2.6 TeV.
The dotted line indicates the vertical flux of atmospheric 
$\mu+\bar{\mu}$ from Ref. 15.}
\end{figure}

The TeV photons 
observed by Whipple collaboration\cite{AGN} may be byproducts of hadronic 
cascades initiated by the protons generated within the AGN accretion 
disk of gas, or in the jets, 
which interact 
with matter or radiation in the AGN disk, 
to produce pions whose decay products include both
photons
and neutrinos.
The structure of the corresponding hadronic cascade is:
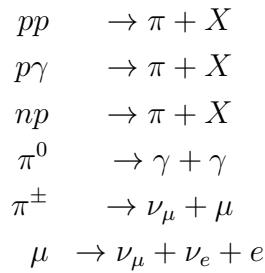
\begin{eqnarray}
p p &\rightarrow \pi  + X\nonumber\\
p \gamma &\rightarrow \pi + X\nonumber\\
n p &\rightarrow \pi  + X\nonumber\\
 \pi^0 &\rightarrow \gamma + \gamma\nonumber\\
 \pi^{\pm} &\rightarrow \nu_\mu + \mu\nonumber\\
 \mu &\rightarrow \nu_\mu + \nu_e + e\nonumber
\end{eqnarray}
If charged and neutral pions are produced in equal proportions and
photons originate in hadronic cascades,
simple counting leads to equal fluxes of photons 
and $\nu_\mu+\bar{\nu}_\mu$. The flux of $\nu_e+\bar{\nu}_e$ equals half
of the flux of $\nu_\mu+\bar{\nu}_\mu$. The observed photon energy spectrum
is a power-law
with\cite{AGN} 
$$ \frac {d N_\gamma}{dE_\gamma} \sim E_\gamma^{-2}$$
for 100 MeV$\,\leq E_\gamma\leq 2$ TeV, and the same for
neutrinos. We have chosen three representative fluxes of neutrinos
from AGN, each corresponding to the diffuse flux integrated 
over all AGNs.  
These fluxes are shown in Figure 1.
The Nellen, Mannheim and Biermann flux\cite{mannheim}  (AGN-NMB), which
comes from assuming that $pp$ collisions are the dominant neutrino source,
is parameterized by:
$${dN_{\nu_{\mu}+\bar\nu_{\mu}}\over dE_{\nu}}=
1.13 \times 10^{-12} (E_{\nu}/{\rm TeV})^{-2}{\rm cm}^{-2}{\rm s}^{-1} 
{\rm sr}^{-1}
{\rm GeV}^{-1}$$
with
the {$\nu_e + \bar \nu_e$ spectrum  assumed to be 
1/2 of $\nu_\mu + \bar \nu_\mu$}.  The 
neutrino luminosity of a source is normalized 
to the observed diffuse 
x-rays and $\gamma$-rays.  
The NMB parameterization is valid for $E_\nu\leq 4\times 10^4$ GeV. In our
calculations described in the next section, we have used this parameterization
up to $E_\nu=10^8$ GeV.
A somewhat different assumption of the luminosity is used by
Szabo and Protheroe\cite{sp}  (AGN-SP) in their extended model of 
neutrino sources,
yielding a higher normalization of $dN/dE_\nu$
at 1 TeV. Above $E_\nu>10^6$ GeV, the AGN-SP follows a steeper power
law, 
$$dN/dE_\nu\sim E^{-3.5}$$
which accounts for the lack of protons at even higher energies required to
produce neutrinos. The Stecker and Salamon flux\cite{stecker}  (AGN-SS)
contains contributions from both $pp$ and $p\gamma$ interactions in the 
accretion disk and has 
a nearly constant value of $dN/dE_\nu$ up
to $E_\nu\sim 10^5$ GeV.

Two models of neutrino fluxes from cosmic ray interactions with the
microwave background\cite{teshima} 
are labeled CR-2 and CR-4 in Figure 1. The fluxes
depend on the redshift of the cosmic ray sources. Maximum redshifts
contributing are $z_{max}=2$ and $z_{max}=4$, respectively.

The electron neutrino plus antineutrino fluxes, to a good approximation,
are equal to half of the fluxes shown in Figure 1.

\bigskip
\noindent{\bf UHE MUON NEUTRINOS}
\medskip

The primary means of detection of muon neutrinos and antineutrinos
is by charged-current conversion into muons and antimuons. The long
range of the muon means that the effective volume of an underground
detector can be significantly larger than the instrumented volume.
For example, a 10 TeV muon produced by a charged-current interaction in
rock will propagate several kilometers in water-equivalent distance units
before its energy is degraded to 1 TeV.

Backgrounds to AGN sources of $\nu_\mu+\bar{\nu}_\mu$ include
atmospheric neutrinos and atmospheric muons.
Muons produced by cosmic ray interactions in the atmosphere
mask astrophysical signals unless detectors are very deep underground,
muon energy thresholds are set very high,
or one observes upward-going muons. 
We evaluate here event rates for upward-going muons produced in
the rock surrounding the detector, for muon energy thresholds
above 1 TeV and 10 TeV.

The neutrino-nucleon cross section comes into the calculation of
the event rate in two ways. The probability of conversion $\nu_\mu
\rightarrow \mu$ is proportional to the $\nu N$ charged current
cross section. In addition, the neutrino flux is attenuated by passage
through the Earth. In the next section we describe our calculation of the
neutrino-isoscalar nucleon ($\nu N$) cross section. The $\nu N$ 
charged-current reaction is the dominant source of neutrino interactions 
except in a very narrow energy window
at the $W$-boson resonance.

\bigskip
\noindent{\bf SMALL-$x$ PARTON DISTRIBUTION FUNCTIONS AND 
$\sigma(\nu N)$}
\medskip

The 
inclusive cross section for $\nu_{\mu} + N \rightarrow
\mu^{-} + X$ is given by
\begin{equation}
\frac {d^2\sigma}{dxdy}=\frac{2G_F^2ME_\nu}{\pi}
\frac{M_W^4}{{(Q^2+M_W^2)}^2} [xq(x,Q^2)+x(1-y)^2\bar q(x,Q^2)],
\end{equation}
where $x=Q^2/2M\nu$, $y=\nu/E_\nu$,
with $-Q^2$ the momentum transfer between the neutrino and muon, and
$\nu$  the lepton energy loss in the lab frame, $\nu=E_\nu -E_\mu$.
$M$ is the mass of the nucleon and $M_W$ is the mass of the $W$-boson, while
the Fermi constant is $G_F=1.16 \times 10^{-5}$ GeV$^{-2}$. Taking 
the target as isoscalar nucleons, in terms of the parton distribution
functions for the proton,
\begin{equation}
q(x,Q^2)=\frac{u_v + d_v}{2} +
\frac{u_s + d_s}{2} +
s_s + b_s
\end{equation}
\begin{equation}
\bar q(x,Q^2)=\frac{u_s + d_s}{2} + c_s + t_s 
\end{equation}
where we have written explicitly valence ($v$) and sea ($s$) distributions.

\begin{figure}[htbp]
\vspace*{6.0 cm}
\caption{Comparison of the light-quark sea at $Q^2=M_W^2$ for various
parton distribution functions.  Of the MRS distributions, D\_ (A') is 
the most (least) singular.  
}
\end{figure}

The general form of the cross section shows that at low energies, where
the four-Fermi approximation is valid,
$\sigma\sim E$. At higher energies, the $W$-boson propagator plays
an important role. The value of $\langle Q^2\rangle $ 
saturates at  $\sim M_W^2$, and 
$x\sim M_W^2/(2ME_\nu y)$ decreases. For neutrino energies above
$10^5$ GeV, the small-$x$ ($x\leq 3\times 10^{-2}$) behavior
of the parton distribution functions becomes important for the
evaluation of the cross section.

\begin{figure}[htbp]
\vspace*{6.0 cm}
\caption{The charged-current cross section for the CTEQ-DIS,
CTEQ-DLA, EHLQ-DLA, MRS A', MRS G and MRS D\_ 
parton distribution functions.  The data point, 
an average of ZEUS and H1, 
is
from Ref. 17.}
\end{figure}

Neutrino charged-current interactions have been measured directly
in laboratory experiments for neutrino energies up to
$E_\nu=300$ GeV\cite{CCFR}. Charged-current $ep$ scattering at HERA, equivalent
to $E_\nu=47.4 $ TeV, can be translated to a value 
of $\sigma(\nu N)$\cite{Cc}.
Recent ZEUS and H1 measurements at HERA\cite{He}
of $F_2^{ep}$ at small-$x$ 
($10^{-4}\leq x \leq 10^{-2}$) and for a large range of $Q^2$, 
4 GeV$^2 \leq Q^2 \leq 1600$ GeV$^2$ have provided valuable information 
about parton densities at small-$x$ and low-$Q^2$.
To evaluate the neutrino-nucleon cross section at ultrahigh
energies, extrapolations beyond the measured regime in
$x$ and $Q^2$ are required.

There are two main theoretical approaches in the evolution
in $Q^2$  of parton
densities: Gribov-Lipatov-Altarelli-Parisi\cite{Gl}
(GLAP) evolution and Balitskii-Fadin-Kuraev-Lipatov\cite{Kl}
(BFKL) evolution. In the GLAP approach, parton distribution
functions are extracted at modest values of $Q^2$ and evolved to
higher scales. The BFKL approach involves a leading $\alpha_s\ln(1/x)$
resummation of soft gluon emissions, which generates a singular behavior
in $x$ at an initial scale $Q_0$,
\begin{equation}
xq_s(x,Q_0^2)\sim x^{-0.5}
\end{equation}
for small $x$,
which persists at higher values of $Q$.
In our extrapolation of the parton distribution functions outside
the measured region, we use GLAP evolution with 
input at $Q_0=1.6$ GeV,
\begin{equation}
xq_s(x,Q_0^2)\sim x^{-\lambda}.
\end{equation}
The value of 
$\lambda$ is determined by fits to deep-inelastic scattering and
hadron-hadron data by the MRS\cite{Ms} and CTEQ\cite{CTEQ} 
Collaborations. The
MRS set A' has $\lambda=0.17$, the MRS set G has 
$\lambda=0.07$  
while the MRS set D\_ has $\lambda=0.5$.  
All of the MRS distribution function are fitted 
using the $\overline{\rm MS}$ factorization scheme.
The CTEQ-DIS, using the deep-inelastic scattering factorization scheme,
has $\lambda=0.33$. These distribution functions are extrapolated using the
power law fit to the distribution functions at $x=10^{-5}$ and
$Q=M_W$.  We have also extrapolated the leading-order CTEQ distributions 
using
the double-log approximation\cite{dla}. 
For reference, the Eichten {\it et al.}\cite{Ehlq}  parton distribution
functions, extrapolated using the double-log approximation, are also
shown. The spread in values for the parton distribution functions
is an indication of the uncertainty in evaluating the
$\nu N $ cross section.

For each of these sets of distribution functions, we have evaluated
the neutrino-nucleon cross section. Figure 3 illustrates the range
of predictions as a function of neutrino energy. 
Also shown is the
average of H1 and ZEUS effective
neutrino nucleon cross sections\cite{Cc}.  
There is excellent agreement among the predictions of
the MRS D\_, G, and A' distributions and the CTEQ3 distributions up
to $E_{\nu}\approx 10^{7}$ GeV.  Above that energy, our DLA
modification of the CTEQ3 distributions gives a lower cross section
than the full CTEQ3 distributions (CTEQ-DIS), as expected from its
less singular behavior as $x \rightarrow 0$.  At the highest energy
displayed, the most singular (MRS D\_) distribution predicts a
significantly higher cross section than the others.  Above about
$10^{6}$ GeV, the EHLQ-DLA distributions yield noticeably smaller
cross sections than the modern distributions.  
Plots similar to Figure 3
for antineutrino-nucleon charged current interactions, as well
as neutral current interactions, can be found in Ref. 6.
For charged current and neutral current interactions, for 10$^{15}$ eV
$\leq E_\nu \leq 10^{21}$ eV, the cross sections follow a simple
power law, for example 
$$\sigma_{{CC}}(\nu N)  =  2.69 \times
        10^{-36} \rm{cm}^{2}\left(\frac{E_{\nu}}
{1\ \rm{GeV}}\right)^{0.402}.  
$$

\newpage
\noindent{\bf NEUTRINO TELESCOPE EVENT RATE}
\medskip

In order to calculate the number of upward-moving muons that can 
be detected with neutrino detectors such as AMANDA, BAIKAL, DUMAND II 
and NESTOR \cite{Detect}, we fold in the neutrino flux and its attenuation
in the Earth with the probability that a neutrino
passing on a detector trajectory 
creates a muon in the rock that traverses the detector.

The attenuation of neutrinos in the Earth 
is described by a shadow
factor $S(E_\nu)$, equivalent to the
effective solid angle for upward muons, normalized to $2\pi$:
\begin{equation}
{d S(E_\nu)\over d\Omega}={1\over 2\pi}
\exp \Bigl( -z(\theta )N_A \sigma_{\nu N}(E_\nu) \Bigr),
\end{equation}
where $N_A=6.022\times 10^{23}$ mol$^{-1}=6.022\times 10^{23}$ cm$^{-3}$
(water equivalent) is Avogadro's number, and $z(\theta)$ is the column
depth of the earth, in water-equivalent units, which depends on zenith 
angle \cite{prem}.
The probability 
that the neutrino with energy $E_\nu$
converts to a muon  is 
proportional to the cross section and depends on the
threshold energy for the muon $E_\mu^{\rm min}$: 
\begin{equation}
P_\mu(E_\nu,E_\mu^{\rm min}) = \sigma_{\rm CC}(E_\nu) N_A \langle
R(E_\nu,E_\mu^{\rm min} )\rangle ,
\end{equation} 
where the average muon range in rock is 
$\langle R\rangle$ \cite{slip}. A more detailed discussion appears in
Ref. 6.

The diffuse flux of AGN neutrinos, 
summed over all AGN sources, is isotropic, so
the event rate is
\begin{eqnarray}
{\rm Rate} = A 
\int dE_\nu P_\mu(E_\nu,E_\mu^{\rm min}) 
S(E_\nu){dN_\nu\over dE_\nu} ,
\end{eqnarray}
given a neutrino spectrum $dN_\nu/dE_\nu$ and detector 
area $A$. As the cross section increases, $P_\mu$ increases, but the
effective solid angle decreases. 

\begin{table}
\caption{
Number of upward $\mu+\bar{\mu}$
per year per steradian for $A=0.1$ km$^2$ and $E_\mu^{\rm min}= 1$ TeV.}
\begin{center}
\begin{tabular}{||l|c|c||}
\hline
Fluxes & EHLQ-DLA & CTEQ-DIS \\ \hline
AGN-SS \cite{stecker} & 82 & 92  \\ \hline
AGN-NMB \cite{mannheim} & 100 & 111  \\ \hline
AGN-SP \cite{sp} & 2660 & 2960 \\ \hline
ATM \cite{volkova}& 126 & 141 \\ \hline
\end{tabular}
\end{center}
\end{table}

\begin{table}
\caption{As in Table 1, but for 
$E_\mu^{\rm min}=10$ TeV.}
\begin{center}
\begin{tabular}{||l|c|c||}
 \hline
Fluxes & EHLQ-DLA & CTEQ-DIS \\ \hline
AGN-SS \cite{stecker} & 46  & 51  \\ \hline
AGN-NMB \cite{mannheim} & 31 & 34  \\ \hline
AGN-SP \cite{sp} & 760 & 843 \\ \hline
ATM \cite{volkova}& 3 & 3  \\ \hline
\end{tabular}
\end{center}
\end{table}

Event rates for upward muons and antimuons
for a detector with $A=0.1$
km$^2$ for $E_\mu^{\rm min}=1$ TeV and $E_\mu^{\rm min}=10$ TeV
are shown in Tables 1 and 2. The CTEQ-DIS distribution functions
are taken as representative of the modern parton distribution
function sets, and compared with the EHLQ-DLA event rate predictions.
The muon range is that of Ref. 26.

The theoretical predictions for ultrahigh-energy neutrinos
from AGNs yield event rates comparable to, or in excess of, the
background rate of atmospheric neutrinos for $E_\mu^{\rm min}=1$ TeV.
The AGN-SP rate is large compared to the AGN-NMB rate because additional 
mechanisms are included.  
Flux limits from the Fr\'ejus experiment
are inconsistent with the SP flux for 1 TeV$< E_\nu <$ 10 TeV 
\cite{frejus}.
The atmospheric neutrino background is greatly reduced by 
requiring a 10 TeV
muon threshold, though AGN induced event rates are reduced as well.
The flatter neutrino spectra
have larger contributions to the event rate
for muon energies away from the
threshold muon energy than the steep atmospheric flux.

We have evaluated the event rates using the other parton distribution
functions shown in Figure 2. Event rate predictions are unchanged
with the other modern parton distributions because all these 
distributions
are in agreement in the energy range $E_\nu \sim 1-100$ TeV.
However, our results for event rates are about 15\% larger than 
for the EHLQ structure functions.  
This is 
due to the fact that EHLQ parton distributions 
were based on 
the CERN-Dortmund-Heidelberg-Saclay measurements of neutrino-nucleon 
structure functions \cite{CCFR} which had 
low normalization of about 15\%.  

\bigskip
\noindent{\bf UHE ELECTRON NEUTRINOS}
\medskip

Finally we consider event rates from electron neutrino and antineutrino
interactions. For $\nu_eN$ (and $\bar{\nu}_eN$) interactions, the
cross sections are identical to the muon neutrino (antineutrino) nucleon
cross sections. 
Because of the rapid energy loss or 
annihilation of electrons and positrons,
it is generally true that only contained-vertex events can be observed.  
Since electron neutrino fluxes are small, 
an extremely large effective volume
is needed to get measurable event rates.  
There is one exceptional case: resonant formation of 
$W^{-}$ in 
$\bar{\nu}_e e$ interactions at 
$E_\nu=6.3$ PeV.  The resonant cross section is larger than the 
$\nu N$ cross section at any energy up to $10^{21}$ eV.  In Fig. 4 
we present neutrino-electron cross sections.  

\begin{figure}[htbp]
\vspace*{10.0 cm}
\caption{Cross sections for neutrino interactions on electron targets.
At low energies, from largest to smallest cross section,
the processes are (i) $\bar{\nu}_{e}e \rightarrow \hbox{ hadrons}$,
 (ii) $\nu_{\mu}e \rightarrow \mu\nu_{e}$, (iii) $\nu_{e}e \rightarrow
 \nu_{e}e$, (iv) $\bar{\nu}_{e}e \rightarrow \bar{\nu}_{\mu}\mu$, (v)
  $\bar{\nu}_{e}e \rightarrow \bar{\nu}_{e}e$, (vi) $\nu_{\mu}e
 \rightarrow \nu_{\mu}e$, (vii) $\bar{\nu}_{\mu}e \rightarrow 
 \bar{\nu}_{\mu}e$.}


\end{figure}

We note that, at the resonance energy, upward-moving electron 
antineutrinos 
do not survive passage through the Earth.  However, the 
contained events 
have better prospects for detection.  
The contained event rate for resonant $W$ production is
\begin{equation}
{\rm Rate} = {10\over 18} V_{\rm eff}  N_A
\int  dE_{\bar{\nu}_{e}}\:\sigma_{\bar{\nu}_{e}e}(E_{\bar{\nu}_{e}}) 
S(E_{\bar{\nu_e}}){dN_{\bar{\nu}_{e}}\over dE_{\bar{\nu_e}}} .
\end{equation}
We show event rates for resonant $W$-boson production in Table 3.  
The background is for events with $E_\nu > 3$ PeV.  

\begin{table}
\caption{
Downward resonance $\bar\nu_e e\rightarrow W^-$ events per 
year per steradian for a detector with effective volume $V_{\rm eff}=1$ 
km$^3$ 
together with the potential downward (upward) background from 
$\nu_\mu$ and $\bar\nu_\mu$ interactions above 3 PeV.}
\begin{center}
\begin{tabular}{||l|c|c||}
\hline
Mode & AGN-SS \cite{stecker} & AGN-SP \cite{sp} \\ \hline
$W\rightarrow \bar{\nu}_\mu \mu$ & 6 & 3 \\ \hline
$W\rightarrow {\rm hadrons}$ & 41 & 19 \\ \hline
$(\nu_\mu,\bar\nu_\mu)$$N$ CC  & 33 (7) & 19 (4) \\ \hline
$(\nu_\mu,\bar\nu_\mu)$$N$ NC  & 13 (3) & 7 (1) \\ \hline
\end{tabular}
\end{center}
\end{table}

From Table 3 we note that a 1 
km$^3$ detector with energy threshold in the PeV range 
would be suitable for 
detecting resonant $\bar\nu_e e \rightarrow W$ events.  
However, the $\nu_\mu N$ background may be difficult to overcome.
By placing the detector a few km underground, one can reduce 
atmospheric-muon 
background, which is 
5 events per year per steradian at the surface of 
the Earth for $E_\mu>3$ PeV.  

\bigskip
\noindent{\bf SUMMARY}
\medskip

In summary, we find that detectors such as 
DUMAND II, AMANDA, BAIKAL 
and NES\-TOR have a very good chance 
of being able to test different models for
neutrino production in the AGNs \cite{review}.    
For $E_{\mu}^{\rm min}=1$ TeV, we find that 
the range of theoretical fluxes leads to event rates of 
900-29,600
upward-moving 
muons/yr/km$^2$/sr
originating from 
the diffuse AGN neutrinos, with the atmospheric background of 1400 
events/yr
/km$^2$/sr. 
For $E_{\mu}^{\rm min}=10$ TeV, 
signal to background ratio becomes even better, 
with signals being on the order of 500-8,400 events/yr/km$^2$/sr,
a factor $\sim$20-300 higher than the background rate.  
For neutrino energies above $3$ PeV there is significant contribution 
to the muon rate due to the $\bar \nu_e$ interaction with electrons, 
due to the $W$-resonance contribution.  We find that 
acoustic detectors with 3 PeV threshold and with 
effective volume of 
0.2 km$^3$, such as DUMAND, would detect 
48 hadronic 
cascades per year 
from 
$W \rightarrow$ hadrons, 7 events from 
$W \rightarrow \mu \bar\nu_{\mu}$ and 
36 events from $\nu_{\mu}$ and 
$\bar \nu_{\mu}$ interactions with virtually no background from 
ATM neutrinos.  
\smallskip
\subsection*{Acknowledgements}
\smallskip
This work was supported in part by DOE grants 
DE-FG03-93ER40792, DE-FG02-85ER40213, and NSF Grant
PHY 95-07688.  
Fermilab is operated by Universities 
Research Association, Inc., under
contract DE-AC02-76CHO3000 with the United States 
Department of Energy.

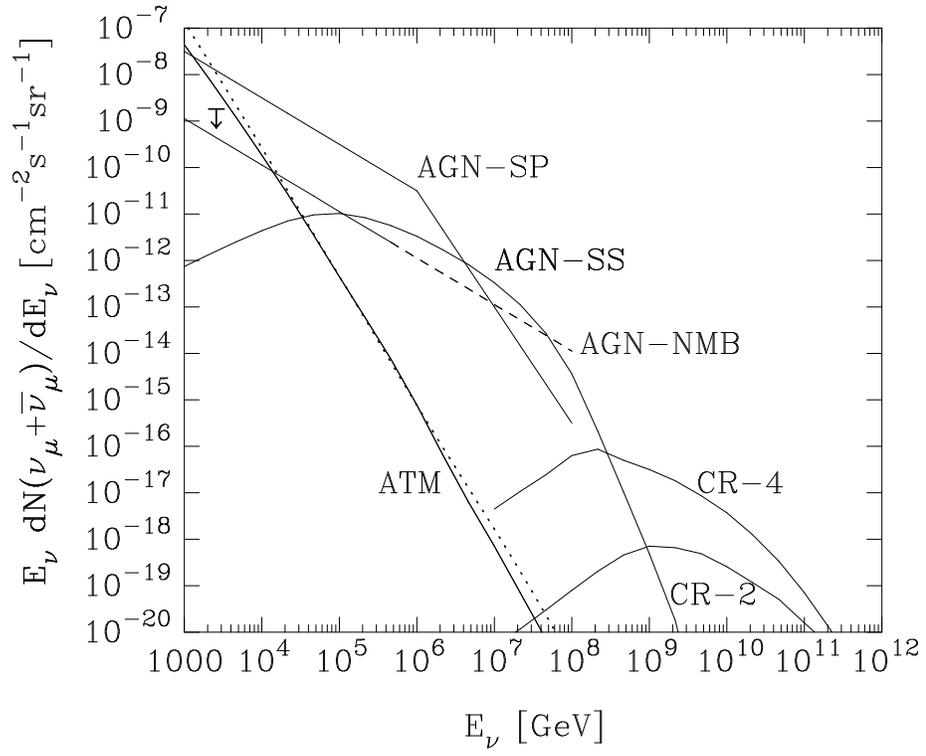
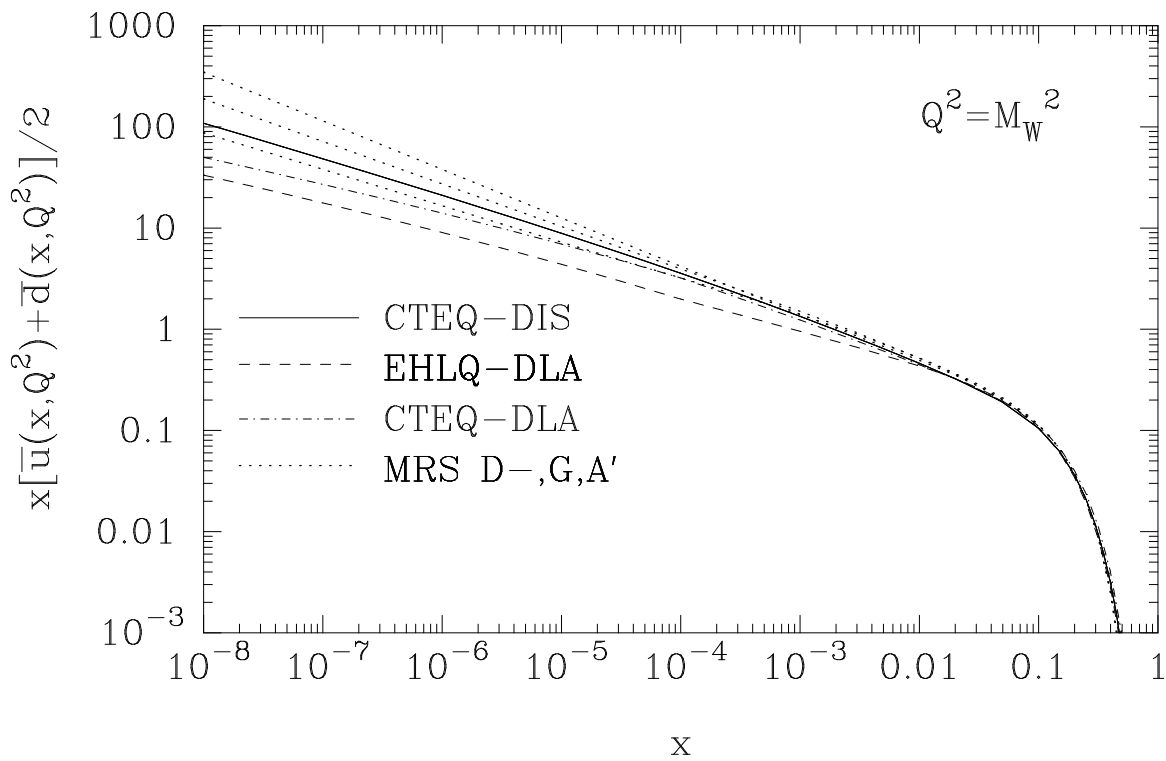
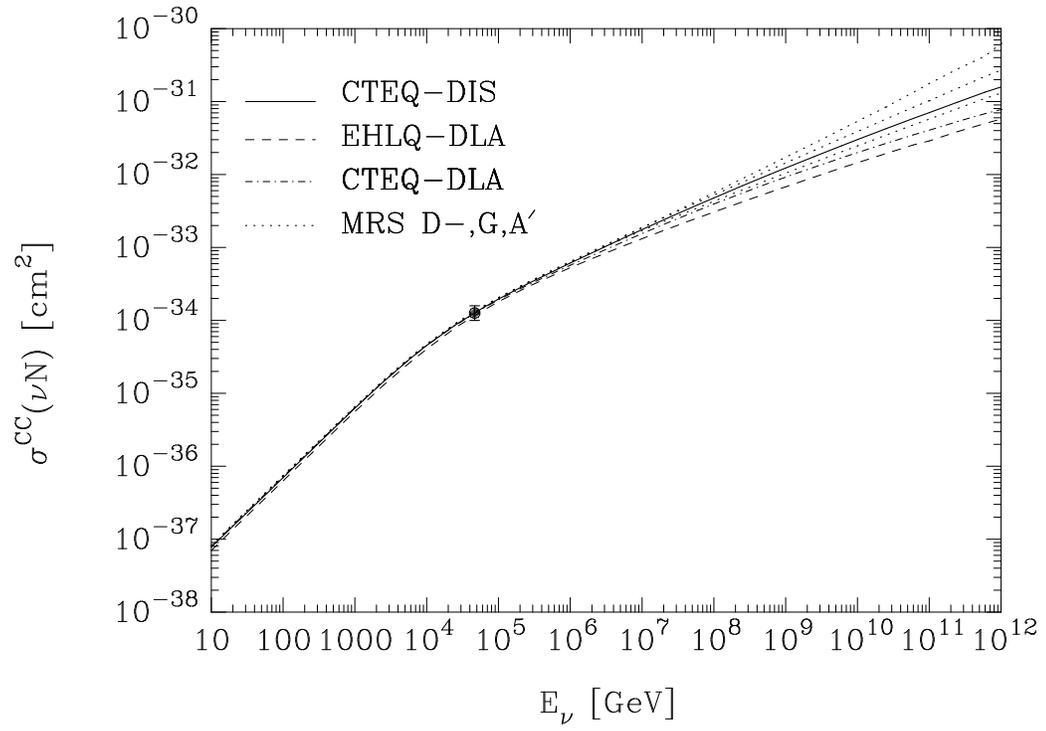
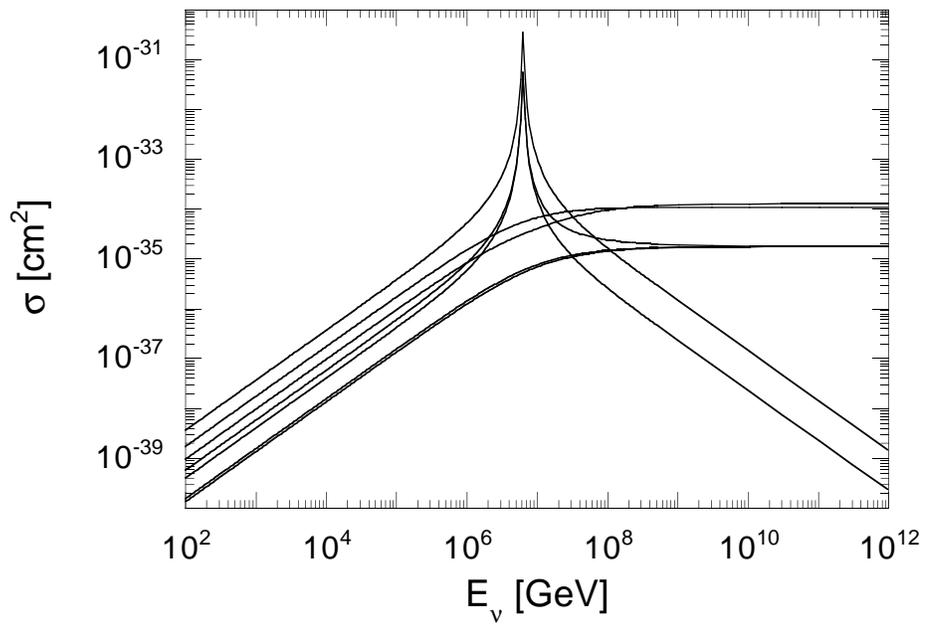
\end{document}